\documentclass[nohyper,11pt,letterpaper]{JHEP3}
\usepackage[dvips]{epsfig}
\usepackage{amsmath}

\newcommand{\be}{\begin{equation}}
\newcommand{\ee}{\end{equation}}
\newcommand{\bea}{\begin{eqnarray}}
\newcommand{\eea}{\end{eqnarray}}
\newcommand{\ba}{\begin{eqnarray}}
\newcommand{\ea}{\end{eqnarray}}
\newcommand{\nn}{\nonumber \\}
\newcommand{\eqn}[1]{(\ref{#1})}
\newcommand{\beq}{\begin{equation}}
\newcommand{\eeq}{\end{equation}}
\newcommand{\beqa}{\begin{eqnarray}}
\newcommand{\eeqa}{\end{eqnarray}}
\newcommand{\beqar}{\begin{eqnarray*}}
\newcommand{\eeqar}{\end{eqnarray*}}

\newcommand{\reef}[1]{(\ref{#1})}

\newcommand{\eg}{{\it e.g.,}\ }
\newcommand{\ie}{{\it i.e.,}\ }

\newcommand{\dd}{\tilde{d}}
\def\nc {N_\mt{c}}
\def\nf {N_\mt{f}}
\def\gym {g_\mt{YM}}
\newcommand{\td}{T_\mt{dec}}
\newcommand{\tf}{T_\mt{fun }}

\newcommand{\mq}{M_\mt{q}}      
\newcommand{\N}{{\cal N}} 
\newcommand{\mbar}{\bar{M}}
\newcommand{\gs}{g_\mt{s}}
\newcommand{\ls}{\ell_\mt{s}}
\newcommand{\ids}{I_\mt{D7}}
\newcommand{\ide}{I_\mt{E}}
\newcommand{\nb}{n_\mt{b}}
\newcommand{\nq}{n_\mt{q}}

\newcommand{\mt}[1]{\textrm{\tiny #1}}
\newcommand{\mc}{M_\mt{c}} 

\def\sac{\, , \,\,\,\,\,}

\newcommand{\labell}[1]{\label{#1}} 

\newcommand{\ra}{\rightarrow}
\newcommand{\overlrarrow}[1]{\vbox{\ialign{##\cr\cr
                  \leftrightarrowfill\crcr\noalign{\kern-1pt\nointerlineskip}
                  $\hfil\displaystyle{#1}\hfil$\crcr}}}
\newcommand{\mub}{\mu_\mt{b}}
\newcommand{\muq}{\mu_\mt{q}}

\newcommand{\nbar}{\bar{\mathcal{N}}}

\title{Holographic phase transitions at finite chemical potential}

\author{David Mateos,$^a$ Shunji Matsuura,$^{b,c}$
Robert C. Myers,$^{b,d}$ and Rowan M. Thomson$^{b,d}$\\
$^a$ Department of Physics, University of California, Santa Barbara,
CA 93106-9530, USA\\
$^b$ Perimeter Institute for Theoretical Physics,
Waterloo, Ontario N2L 2Y5, Canada \\
$^c$ Department of Physics, University of Tokyo,
7-3-1 Hongo, Bunkyoku, Tokyo\\
\ \ 113-0033, Japan\\
$^d$ Department of Physics and Astronomy, University of Waterloo,
Waterloo, Ontario\\
\ \  N2L 3G1, Canada \\
\\E-mail: \email{dmateos@physics.ucsb.edu, smatsuura@perimeterinstitute.ca,
rmyers@perimeterinstitute.ca, rthomson@perimeterinstitute.ca }}

\abstract{Recently holographic techniques have been used to study
the thermal properties of ${\cal N}=2$ super-Yang-Mills theory, with
gauge group $SU(\nc)$ and coupled to $\nf \ll \nc$ flavours of
fundamental matter, at large $\nc$ and large 't Hooft coupling. Here
we consider the phase diagram as a function of temperature and
baryon chemical potential $\mub$. For fixed $\mub < \nc\,\mq$ there
is a line of first order thermal phase transitions separating a
region with vanishing baryon density and one with nonzero density.
For fixed $\mub> \nc\,\mq$ there is no phase transition as a
function of the temperature and the baryon density is always
nonzero. We also compare the present results for the grand canonical
ensemble with those for canonical ensemble in which the baryon
density is held fixed \cite{findens}.}


\preprint{arXiv: [hep-th]}

\begin{document}


\section{Introduction}

A large class of strongly coupled gauge theories can be studied
using the gauge/gravity duality \cite{juan,bigRev}. In this context,
the dynamics of a small number of flavours $\nf \ll \nc$ of
fundamental matter in the gauge theory can be described by the
dynamics of $\nf$ D-branes in the appropriate dual geometry. At
sufficiently high temperatures, for which the gauge theory is in a
deconfined or plasma phase,\footnote{The theory may or may not
undergo a phase transition to a confined phase at low temperatures.}
this geometry contains a black hole \cite{witten}. Since the
condition $\nf \ll \nc$ ensures that the D-branes only perturb the
background slightly, the dynamics of the fundamental matter in the
deconfined phase corresponds to the dynamics of D-brane probes in a
black hole background.

The thermal properties of $\nf$ flavours of fundamental matter in
$SU(\nc)$ super-Yang-Mills (SYM) have been extensively studied using
this holographic framework \cite{johanna,prl,long,visco,recent}. In
these systems, the fundamental matter generically undergoes a first
order phase transition at some critical temperature $\tf$.
The low-temperature phase of the theory is described by `Minkowski
embeddings' of the probe branes (see fig.~\ref{embeddings}) in which
the branes sit entirely outside the black hole \cite{prl,long}. In
this phase, the meson spectrum is discrete and exhibits a mass gap.
Above the critical temperature $T_\mt{fun}$, the branes fall through the
horizon and the gravity system is characterised by these `black
hole' embeddings.  In this phase, the meson spectrum is continuous
and gapless \cite{long,spectre,hoyos}.  Thus, this large-$\nc$,
strong coupling phase transition is associated with the melting or
dissociation of mesons. As recently illustrated in related models
\cite{recent8}, we emphasize that in theories that undergo a
confinement/deconfinement phase transition at some temperature
$\td < T_\mt{fun}$, the phase transition for fundamental matter just described
occurs as a separate phase transition. Instead, if $\tf < \td$, both
phase transitions take place simultaneously. In the former case,
mesonic states remain bound for a range of temperatures $\td < T <
\tf$ despite the fact that the theory is in a deconfined phase.
\FIGURE{
 \includegraphics[width=1 \textwidth]{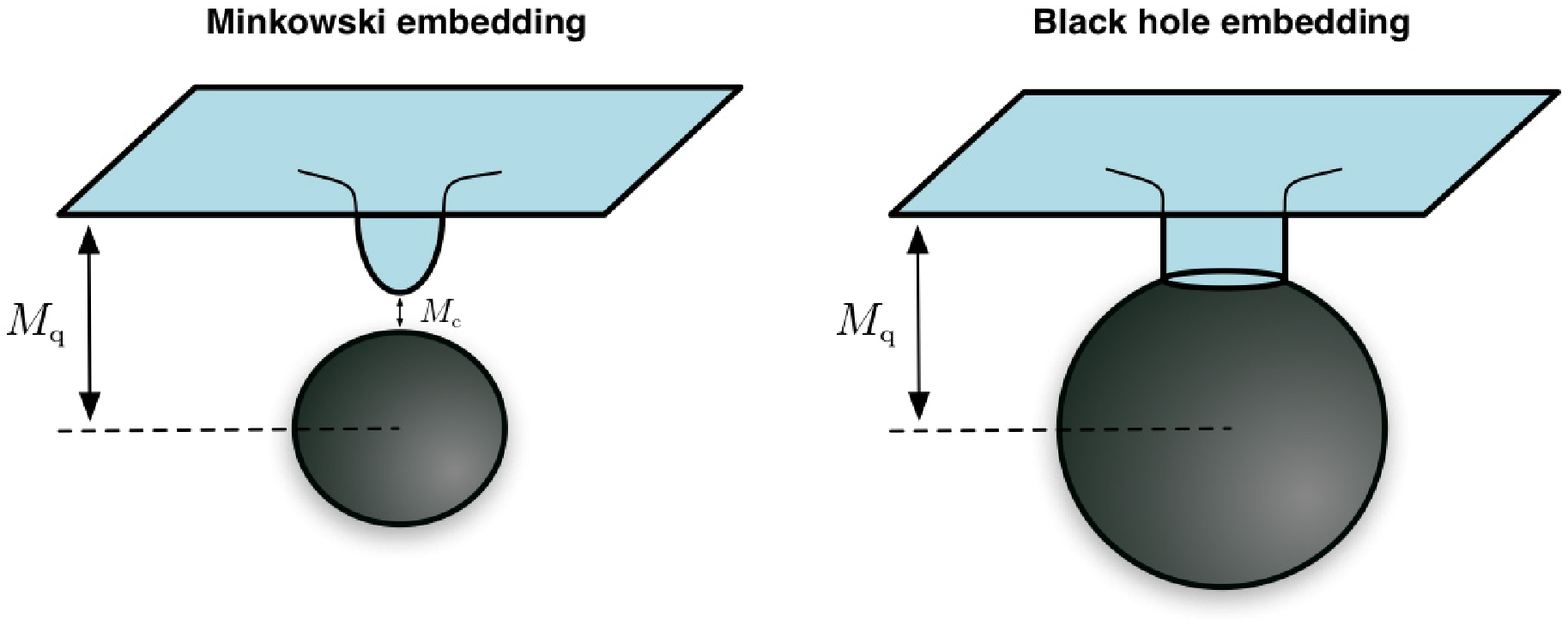}
\caption{Possible D-brane embeddings in the black hole geometry.}
\label{embeddings}}

This physics is in qualitative agreement with that of QCD.   Studies
from lattice QCD \cite{lattice} suggest that some mesons survive the
deconfinement phase transition at $\td \sim 175 $ MeV and remain as
relatively well-defined resonances up to temperatures of $2-3\,
\td$. It is therefore interesting to consider how this physics is
modified in the presence of a chemical potential $\mub$ or nonzero
density $\nb$ for baryon number. In the holographic framework, the
introduction of a chemical potential $\mub$ or nonzero density $\nb$
for baryon number corresponds to turning on the diagonal $U(1)$
gauge field on the D-branes \cite{findens}.

The phase diagram for the gauge theory at constant baryon number
density $\nb$ was studied in \cite{findens}. It was found that, for
any nonzero value of the baryon number density $\nb$, Minkowski
embeddings are physically inconsistent. Simply stated, a nonzero
density of baryons (or quarks) is dual to a worldvolume electric
field, which in turn translates into a finite number of
(fundamental) strings dissolved in the probe D-branes.  Since the
strings cannot simply terminate, it is not possible for the branes
to close off smoothly above the horizon. Fortunately, for $\nb \neq
0$ there exist black hole embeddings where the D-branes intersect
the horizon for all values of the temperature. This is in contrast
with the case $\nb=0$, in which black hole embeddings do not exist
below a certain temperature and Minkowski embeddings are required to
describe the system \cite{prl,long}. Focussing on the D3/D7 system,
\ie $\nf$ D7-brane probes in the black near-horizon geometry of
$\nc$ D3-branes, it was shown in ref.~\cite{findens} that the
physics is essentially continuous around $\nb=0$. For small $\nb$,
the black hole embeddings mimic the behaviour of both the Minkowski
and black hole branches of the $\nb=0$ system, and a first order
phase transition occurs between two different black hole embeddings
-- see ref.~\cite{findens} for details.  In a region with small
$\nb$, the $(T,\nb)$ phase diagram exhibits a line of thermal phase
transitions which terminates at a critical point (at finite $\nb$).
However, it must be noted that the configurations below the phase
transition were found to be unstable suggesting the existence of a
new, perhaps inhomogeneous, phase. Though the focus in
ref.~\cite{findens} was on the D3/D7 system, general arguments
presented there suggest that the same physics occurs in other Dp/Dq
systems, corresponding to gauge theories in different dimensions
\cite{shunji}.

The above study of the thermodynamics with fixed $\nb$ was performed
in the canonical ensemble.  Of course, the thermodynamics can also
be studied in the grand canonical ensemble, in which the chemical
potential $\mub$ is kept fixed. This is the focus of this paper. As
described above and clearly elucidated in \cite{findens}, the
restriction to black hole embeddings only applies when $\nb$ is
fixed and nonvanishing. Hence, as we will describe, Minkowski
embeddings play an essential role in describing the grand canonical
ensemble. Again we consider only the D3/D7 brane system in the
following but expect similar results for other Dp/Dq brane systems
\cite{shunji}.

Our main result is the phase diagram displayed in fig.~\ref{phaseDiag}, where
\be
\muq = \frac{\mub}{\nc}
\ee
is the quark chemical potential and $\mbar \propto \mq$
is a mass scale defined in eqn.~\reef{mbar} below.
\FIGURE{
\begin{tabular}{cc}
 \includegraphics[width=0.5 \textwidth]{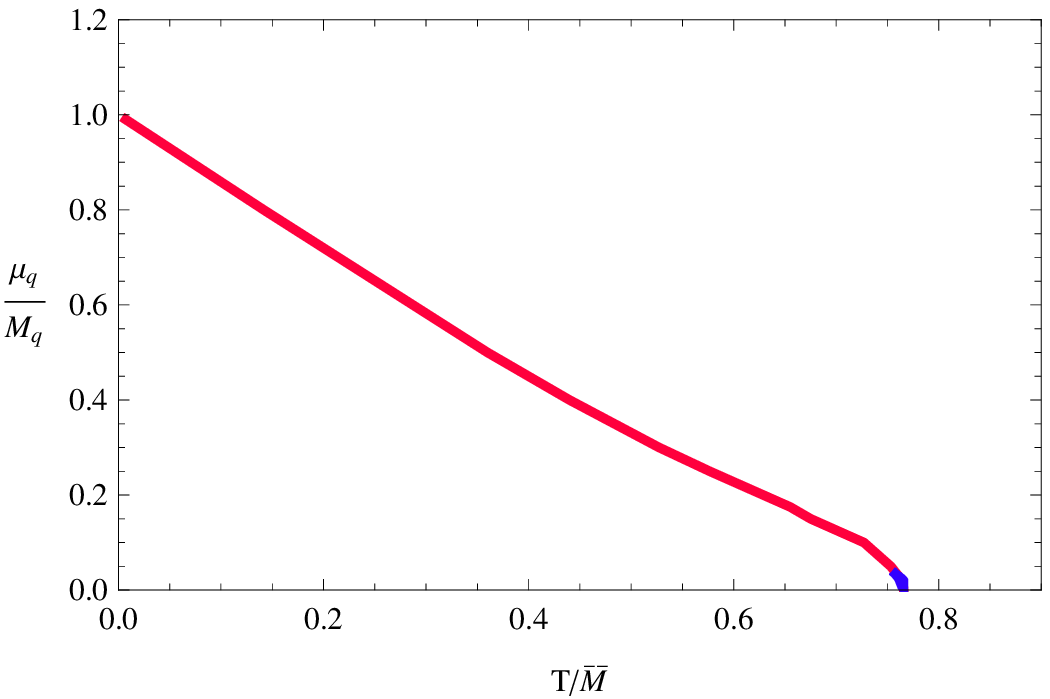}
 \put(-170,50){$\nq=0$}
 \put(-70,90){$\nq \neq 0$} &
 \includegraphics[width=0.5 \textwidth]{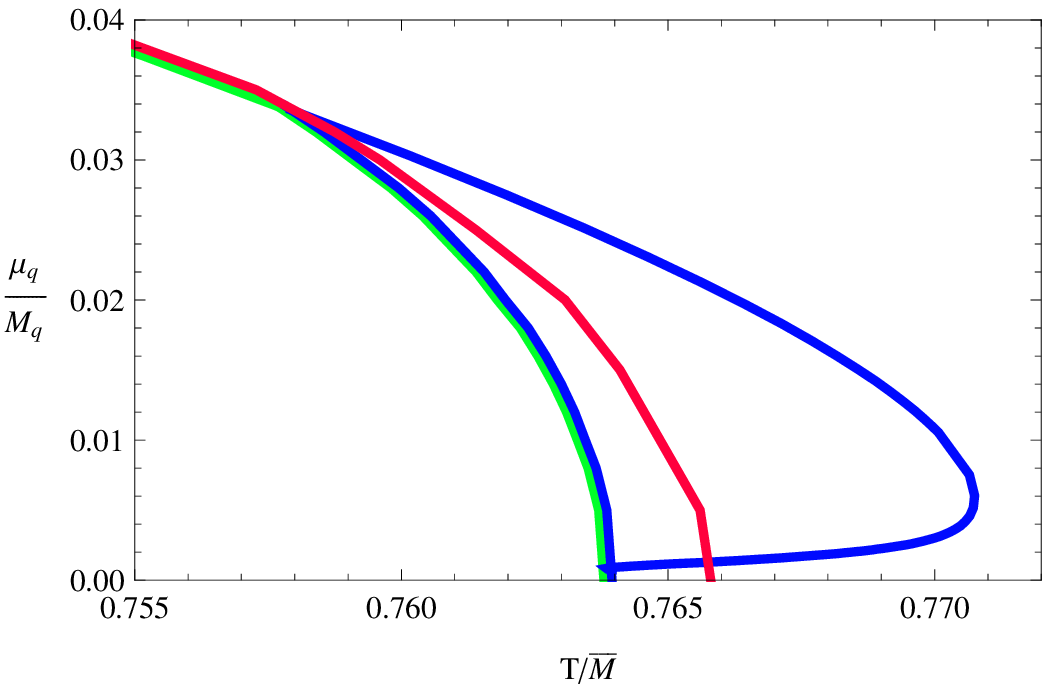} \\
 (a) & (b)
 \end{tabular}
\caption{Phase diagram: Quark chemical potential $\muq/\mq$ versus
temperature $T/\mbar$. The red line separates the phase of Minkowski
embeddings (small temperatures, small $\muq/\mq$) from black hole
embeddings. Figure (b) zooms in on the region near the end of this
line and also depicts the boundary of the region accessed by the
black hole embeddings (green) and a small region (enclosed by the
blue curve) where more than one black hole embedding is available
for a given value of $\muq$ and $T$. The separation of the red and
green curves would not be resolved on the scale of Figure (a).}
\label{phaseDiag} }
As we show in section \ref{framework}, for fixed $\muq < \mq$ black
hole embeddings do not exist below a certain temperature.  In
contrast, for $\muq > \mq$, black hole embeddings exist for all
temperatures. On the other hand, Minkowski embeddings with a
vanishing density of baryons, $\nb=0$, exist for all $\muq$ up to a
temperature $T \sim T_\mt{fun}$. Our study in section \ref{thermo} of the
free energy of the system as the temperature increases reveals that
for $\muq < \mq$ there is a first order phase transition from a
Minkowski embedding to a black hole embedding. In the field theory
this is a transition characterised by the condensation of charge,
namely by a jump from a phase with $\nq=0$ to a phase with
$\nq \neq 0$. The meson spectrum across this transition also changes: In the
low-temperature Minkowski phase the meson  spectrum is discrete and
possesses  a mass gap, whereas in the high-temperature black hole
phase it is continuous and gapless. For larger chemical potential,
$\muq > \mq$, there is no phase transition as a function of the
temperature.

As described above, for $\muq < \mq$ black hole embeddings do not
exist below a certain temperature. The boundary of the portion of
the phase diagram accessed by these embeddings essentially coincides
with the line of phase transitions shown in fig.~\ref{phaseDiag}a.
However, at higher resolution in fig.~\ref{phaseDiag}b, one can see
these two boundaries deviate in a small region near $\muq\sim0$.
This figure also illustrates a small region of the phase diagram
where more than one black hole embedding is available for a given
value of $\muq$ and $T$. In particular, the latter contains the
black hole embeddings that are thermodynamically unstable. As will be
discussed below, this `multi-valued' region of black hole embeddings
was central to the 
analysis of the phase transition studied in
\cite{findens} for the canonical ensemble with fixed $\nq$. In
section \ref{discuss}, we discuss the reconciliation of these
previous results with those found here.

While this paper was in the final stages of preparation, three
related papers \cite{korea,japan,karch3} appeared which have
considerable overlap with our present investigation.

\section{Holographic framework}\label{framework}

Following \cite{prl,long}, the metric for $\nc$ black D3-branes in
the decoupling limit may be written as\footnote{This metric is
related to the standard presentation through the coordinate
transformation $\left( u_0 \rho \right)^2 = u^2 +
\sqrt{u^4-u_0^4}$.}
\beq ds^2 = \frac{1}{2} \left(\frac{u_0 \rho}{L}\right)^2 \left[-
\frac{f^2}{\tilde f} dt^2 + \tilde{f} d\vec{x}^{\,2} 
\right]
 + \frac{L^2}{\rho^2}\left[ d\rho^2 +\rho^2 d\Omega_{\it 5}^2
  \right] \,,
  \labell{geom}
\eeq
where $\rho$ is a dimensionless coordinate and
\be f(\rho)= 1- \frac{1}{\rho^4} \sac
\tilde{f}(\rho)=1+\frac{1}{\rho^4} \sac L^4 = 4 \pi \gs \nc \ls^4
\,. \ee
This metric possesses a horizon at $\rho=1$ with temperature \be T =
\frac{u_0}{\pi L^2} \,, \ee which is identified with the temperature
of the dual ${\cal N}=4$ SYM theory. In turn, the coordinates
$\{t,\vec{x} 
\}$ are identified with the coordinates of the
gauge theory. The string coupling constant is related to the SYM 't
Hooft coupling constant through
\be \lambda = \gym^2 \nc =  2
\pi \gs \nc \,. \ee
In addition to the metric above, the D3-brane solution is
characterised by a constant dilaton and a Ramond-Ramond field given
by
\be C_{0123}=-\frac{\left( u_0 \rho\right)^4 \tilde{f}^2}{4L^4}
\,. \ee

Introducing $\nf$ D7-branes into the  geometry above corresponds to
coupling $\nf$ fundamental hypermultiplets to the original SYM
theory \cite{flavour}. Before the decoupling limit, the branes are
oriented as
\begin{equation}
\begin{array}{ccccccccccc}
   & 0 & 1 & 2 & 3 & 4& 5 & 6 & 7 & 8 & 9\\
\mbox{D3:} & \times & \times & \times & \times & & &  &  & & \\
\mbox{D7:} & \times & \times & \times & \times & \times  & \times
& \times & \times &  &   \\
\end{array}
\labell{D3D7}
\end{equation}
This configuration is supersymmetric at zero temperature, which
ensures stability of the system. After the decoupling limit, the D7
branes wrap an $S^3$ of possibly varying radius inside the $S^5$ of
the background geometry. Writing
\beq
d\Omega_{\it 5}^2 = d\theta^2 +
\sin^2 \theta d\Omega_{\it 3}^2 +\cos^2 \theta d\phi^2
\eeq
and defining $\chi = \cos \theta$, the embedding of the D7-branes is described by
$\phi=0, \, \chi=\chi(\rho)$.

The derivation of the equations of motion for the D7-branes'
embedding $\chi(\rho)$ and the gauge field on their worldvolume
$A_t$ was discussed in \cite{findens} and we review a few salient
details here. The (Dirac-Born-Infeld) action of the D7-branes is
\beq \ids=-\nf T_\mt{D7} \int dt\, d^3\!x \, d\rho \, d\Omega_{\it
3} \frac{\left( u_0 \rho \right)^3}{4} f \tilde{f} (1-\chi^2)
\sqrt{1-\chi^2+\rho^2 \dot{\chi} ^2 - \frac{2
\tilde{f}}{f^2}(1-\chi^2) \dot{\tilde{A_t}}^2} \,, \labell{action}
\eeq where the dot denotes differentiation with respect to $\rho$
and we have introduced a dimensionless gauge field \be \tilde{A}_t =
\frac{2 \pi \ls^2}{u_0} A_t \,. \ee Note that $\dot{\tilde{A_t}}$
corresponds to a radial electric field. The corresponding equation
of motion (eqn. $(2.11)$ in \cite{findens}) dictates that
asymptotically \beq A_t = \muq - \frac{u_0}{2\pi \ls^2}
\frac{\dd}{\rho^2} + \cdots \,, \labell{at} \eeq where the constant
$\muq$ is the quark chemical potential and $\tilde{d}$ is a
dimensionless constant related to the vacuum expectation value of
the quark number density operator through
\beq \nq = \frac{1}{2^{5/2}}\, \nf \nc \sqrt{\lambda}\, T^3\, \dd\,.
\labell{deff1} \eeq
The equation of motion for $\chi$ (eqn.~$(2.17)$ in \cite{findens})
implies  that the branes' profile behaves asymptotically as
\beq
\chi = \frac{m}{\rho}+\frac{c}{\rho^3} + \cdots \,, \labell{asymp}
\eeq
where the dimensionless constants $m$ and $c$ are proportional
to the quark mass and condensate, respectively \cite{prl,long}. In
particular
\be m=\frac{\mbar}{T} \,,\labell{deff2}  \ee
where the scale
\be \mbar = \frac{2\mq}{\sqrt{\lambda}} = \frac{M_\mt{gap}}{2\pi}
\labell{mbar} \eeq
is (up to a factor) the meson mass gap $M_\mt{gap}$ in the D3/D7
brane theory at zero temperature \cite{us-meson}.

Let us now consider what types of embeddings are possible if $\muq
\neq 0$. As shown in \cite{prl,long}, Minkowski embeddings only
exist for $T /\mbar \lesssim 0.77$, \ie for sufficiently low
temperatures or sufficiently large quark masses.  Moreover, these
embeddings are physically consistent only if $\nq=0$ \cite{findens}.
However, the chemical potential can take any value, since a
Minkowski embedding with $A_t = \muq$ is a solution of the equations
of motion if the same embedding is a solution with $A_t =0$. Thus
Minkowski embeddings describe a phase with a possibly nonzero
chemical potential but vanishing charge density. We therefore expect
(and we will confirm below) that these embeddings are
thermodynamically preferred if $\muq \lesssim \mc$, where $\mc$ is
the constituent or thermal quark mass at a given temperature. As
shown in fig.~\ref{embeddings}, this mass is proportional to the
distance between the tip of the D7-branes and the horizon, in
contrast to the bare quark mass $\mq$, which is proportional to the
asymptotic distance between the D7-branes and the D3-branes
\cite{long,seat}.

If $\nq \neq 0$ then only black hole embeddings  are physically
consistent, and these exist for all values of $T / \mbar$.
Regularity requires the gauge field to vanish at the horizon, so
the chemical potential can be expressed in terms of $\dd$, $T$ and
$\mq$ by integrating Gauss' law from the horizon to the boundary as
\cite{findens}
\beq
\muq  = \frac{u_0}{2\pi \ell_s^2} \, 2\tilde{d} \int_1^\infty d\rho\,
\frac{f\sqrt{1-\chi^2+\rho^2\dot{\chi}^2}}
{\sqrt{\tilde{f}(1-\chi^2) \left[ \rho^6 \tilde{f}^3 (1-\chi^2)^3 + 8\tilde{d}^2 \right]}} \,.
\labell{moo}
\eeq
A key observation now is the fact that, for sufficiently small
$T/\mbar$, $\muq$ approaches a nonzero value in the limit $\nq \ra
0$. This is illustrated in fig.~\ref{muVsT}, which displays
$\muq/\mq$ as a function of $T/\mbar$ for decreasing values of $\dd$
(from top curve to bottom curve) all the way down to $\dd =
10^{-4}$.
\FIGURE{
 \includegraphics[width=0.75 \textwidth]{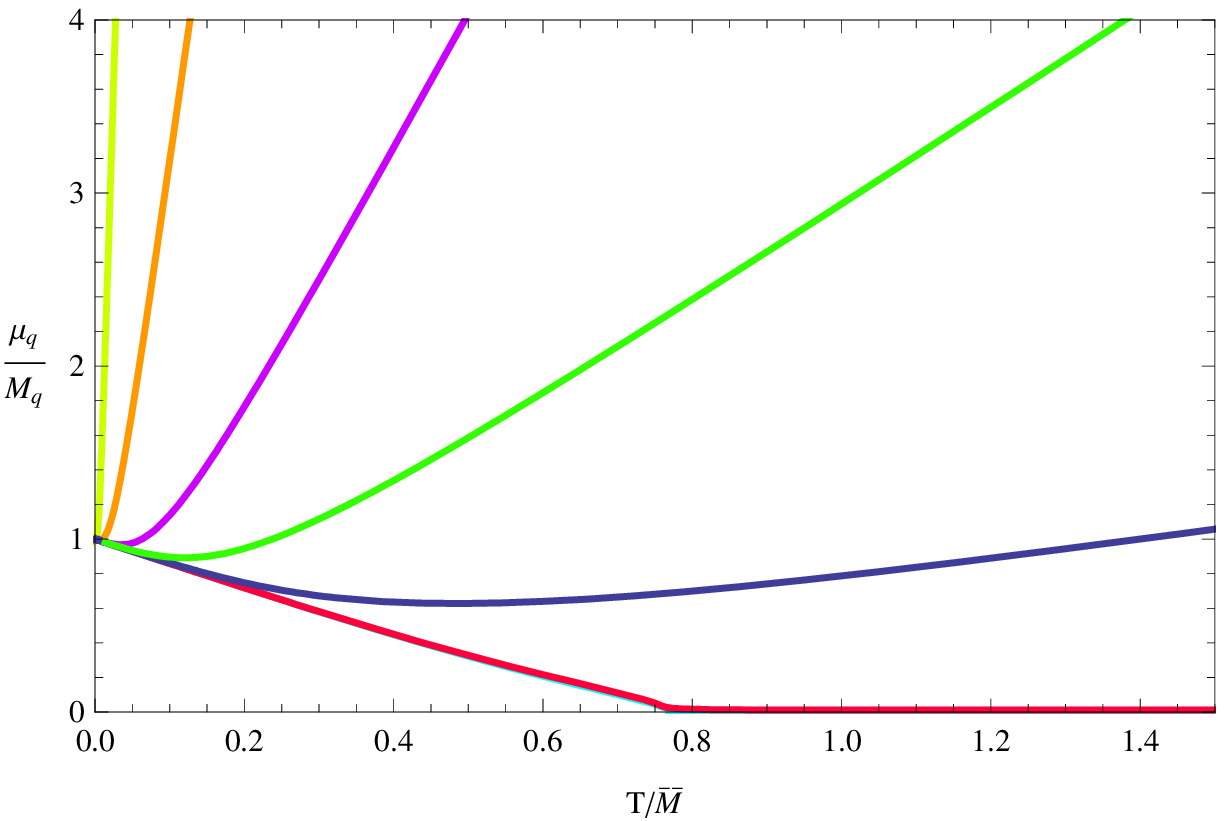}
\caption{Chemical potential $\muq/\mq$ versus  $T/\mbar$ for various
values of $\tilde{d}$, increasing from bottom up: $\tilde{d} =
10^{-4}, \,10^{-2},\,1,\, 10, \,104,\, 4298,\, 380315$. The
$\tilde{d} = 10^{-4}, \,10^{-2}$ lines are virtually coincident.}
\label{muVsT} }
Taking $\tilde{d}$ to be even smaller than this
value does not yield values of $\muq$ appreciably smaller on the
scale shown in the figure. In other words, black hole embeddings
only cover the region above the bottom curve in fig.~\ref{muVsT}.
This observation is important because it means that Minkowski
embeddings  are necessary to describe the region below this curve.

The curves in fig.~\ref{muVsT} were obtained by numerical
integration of eq.~\eqn{moo}. However,  analytic results can be
obtained in the limiting cases of large and small $T/\mbar$. In the
limit $T/\mbar \ra 0$ with fixed $\dd$, it was shown in
\cite{findens} that $\muq \ra \mq$ regardless of the value of $\dd$.
This is reflected in fig.~\ref{muVsT} by the fact that all curves
meet on the vertical axis at $\muq/\mq =1$. In the opposite limit,
$\mbar/T \ra 0$, the embedding becomes simply $\chi(\rho) =0$.
Substituting this into \eqn{moo} one obtains an expression that can
be easily integrated. In particular, one easily verifies in this
case that
\be \muq\simeq\frac{u_0\,\dd}{4\pi\ls^2}
+O(\dd^3)=\frac{2}{\nf\,\nc}\,\frac{n_q}{T^2}+O(n_q^3) \ee
for small $\dd$, and consequently $\muq \ra 0$ as $\dd \ra 0$.

To summarise, we conclude that the region of small $\muq/\mq$ and
small $T/\mbar$ (the triangular region below the bottom line in the
lower left corner of fig.~\ref{muVsT}) must be described by
Minkowski embeddings. Similarly, the region with $T/\mbar \gtrsim
0.77$ must be described by black hole embeddings. In order to decide
what the favoured embedding is in the rest of the parameter space,
\ie in the region in which both Minkowski and black hole embeddings
exist, we must determine which embedding minimises the free energy
of the system.

\section{Thermodynamics}\label{thermo}
From the previous section it is clear that both Minkowski and black
hole embeddings play a role in the grand canonical ensemble.  As
both embeddings exist for certain values of temperature and chemical
potential, it is necessary to study the thermodynamics to determine
the physically preferred configuration. To this end, we use the
standard technique of Wick rotating \cite{hawk} the time direction.
Then, the leading contribution to the free energy is determined by
evaluating the Euclidean action.  As we are interested in the
behaviour of the fundamental matter, we only consider the action for
the D7-branes.

Of course, the Euclidean action for the D7-branes is given by
Wick-rotating the action in \reef{action}.  Evaluating this action
leads to a formally divergent result. However, the divergences are
removed by introducing a finite-radius ultraviolet cut-off and a set
of boundary counterterms to renormalise the action \cite{karch1}.
This holographic renormalisation was discussed in detail in
\cite{findens,prl,long} and the result, from ref.~\cite{findens}, is
\beq \frac{\ide}{\N} = G(m) - \frac{1}{4}
\left[(\rho_\mt{min}^2-m^2)^2-4mc \right] \,, \labell{acta} \eeq
where $G(m)$ is the convergent integral
\beq G(m) = \int_{\rho_\mt{min}}^{\infty} d\rho \left( \rho^3 f
\tilde{f} (1-\chi^2) \sqrt{1-\chi^2+\rho^2 {\dot{\chi}}^2 - \frac{2
\tilde{f}}{f^2} (1-\chi^2) \dot{\tilde{A_t}}^2} - \rho^3 + m^2 \rho
\right) \,, \eeq
${\cal N}$ is a normalisation constant given by
\be\N=\frac{1}{32} \lambda \nf \nc T^3 \,, \ee
and $\rho_\mt{min}$ is the minimum value of $\rho$ for the embedding
of interest.

This Euclidean action of the D7-branes is identified with their
Gibbs free energy $W(T,\muq)$ via $W=T I_\mt{E}$ \cite{findens}. Of
course, the thermodynamically preferred D7-brane embedding is that
which minimises the free energy and hence we evaluated the free
energy as a function of temperature for fixed $\muq/\mq$.
Representative plots of the free energy (scaled by $\nbar=\lambda
\nf \nc \mbar^4/32$) versus temperature are provided in
figs.~\ref{freeEnVsT} and \ref{freeEnVsTbig}. The results for
$\muq/\mq =0.0108$ (fig.~\ref{freeEnVsT}) are typical for small
$\muq/\mq$, displaying the classic `swallow tail' shape and closely
resembling the $\muq =0$ results (shown in fig.~$5$ of \cite{long}).
For larger $\muq/\mq$, the `swallow tail' shape disappears -- see
fig.~\ref{freeEnVsTbig}. For $\muq/\mq<1$, the two branches still
cross at some temperature and this, of course, determines the
temperature of the phase transition -- see fig.~\ref{freeEnVsTbig}a.
As $\muq/\mq$ increases towards 1, the temperature of the phase
transition decreases.  The results in fig.~\ref{freeEnVsTbig}b are
typical for large $\muq/\mq$:  The two branches never cross and the
free energy is always lower on the black hole branch.
\FIGURE{
\begin{tabular}{cc}
 \includegraphics[width=0.5 \textwidth]{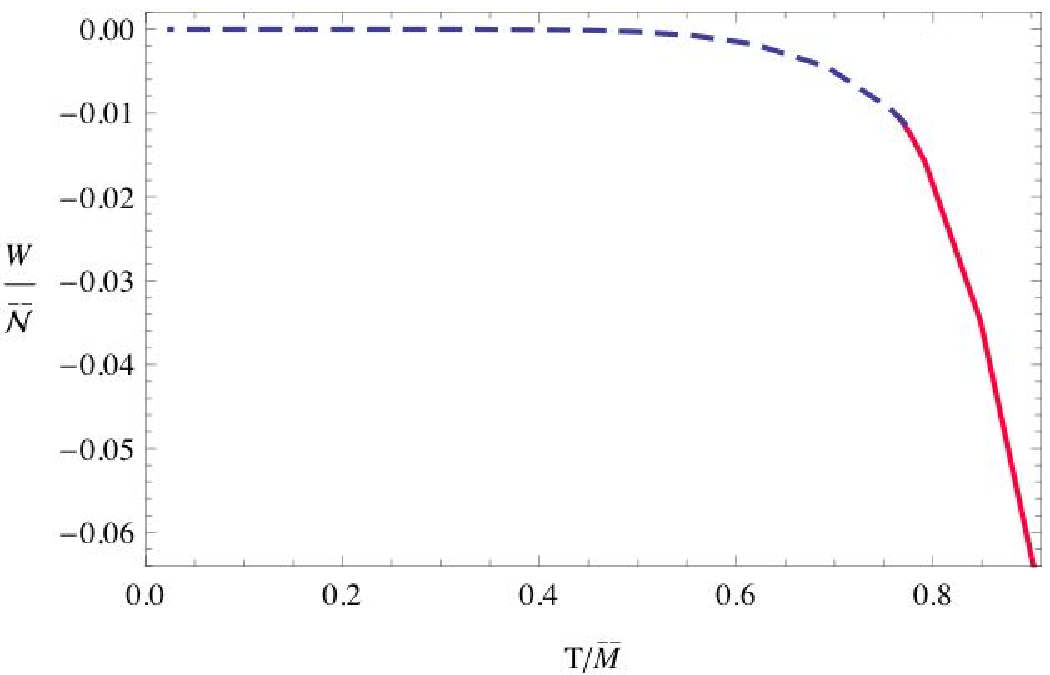}&
 \includegraphics[width=0.5 \textwidth]{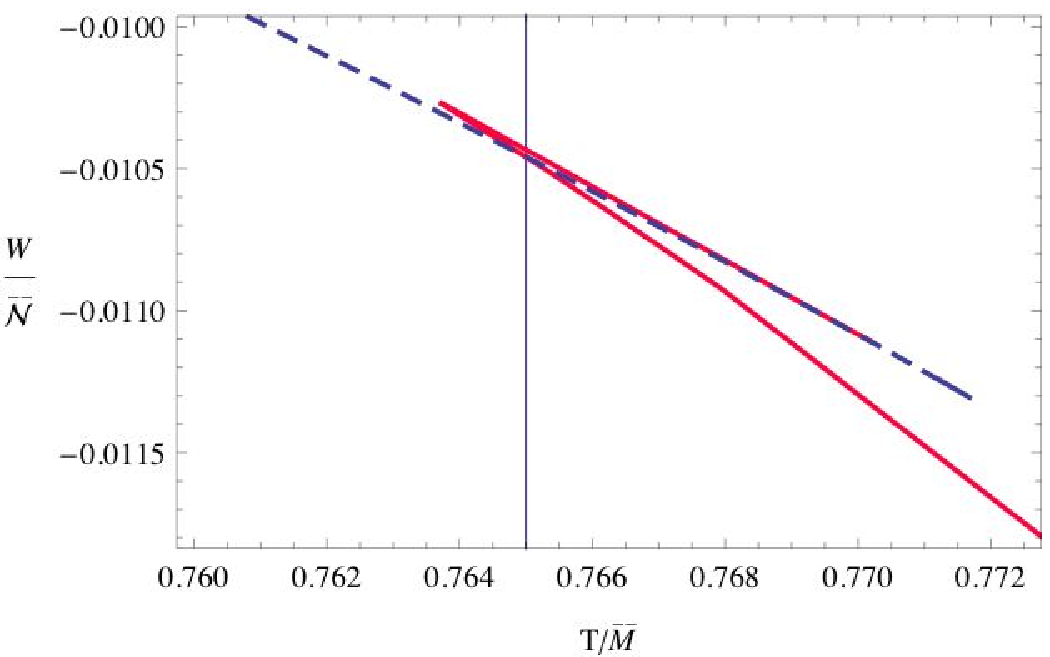}
 \end{tabular}
\caption{Free energy versus temperature for $\muq/\mq=0.0108187$. The
blue dotted line (red solid) represents the Minkowski (black hole)
branch.  The vertical line marks the temperature of the phase
transition.} \label{freeEnVsT}}
\FIGURE{
\begin{tabular}{cc}
  \includegraphics[width=0.5 \textwidth]{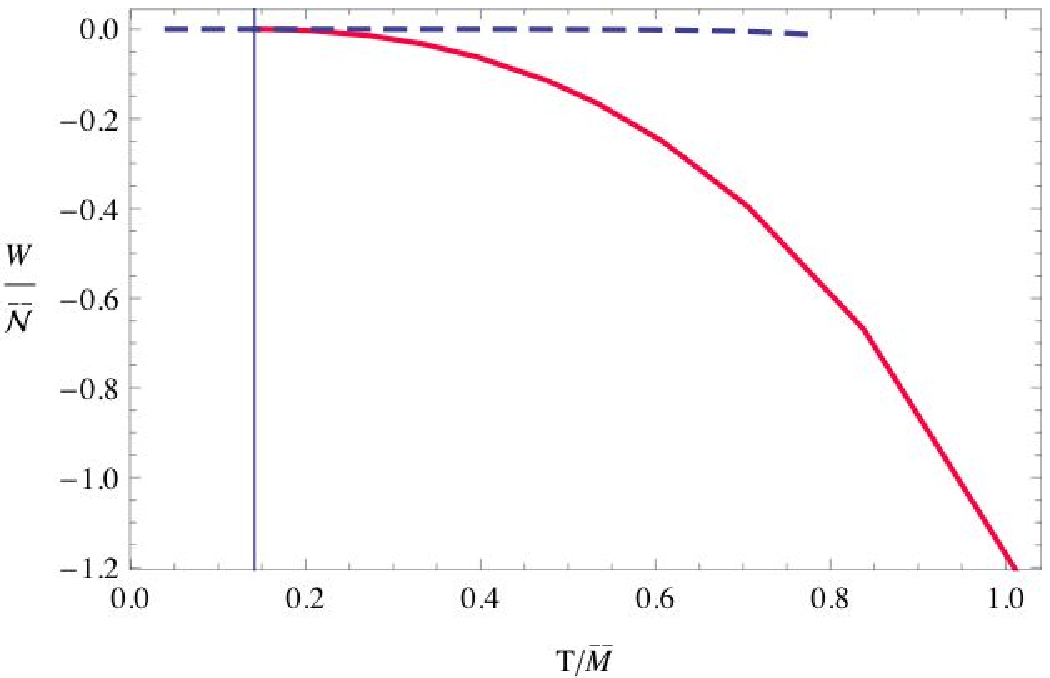}&
 \includegraphics[width=0.5 \textwidth]{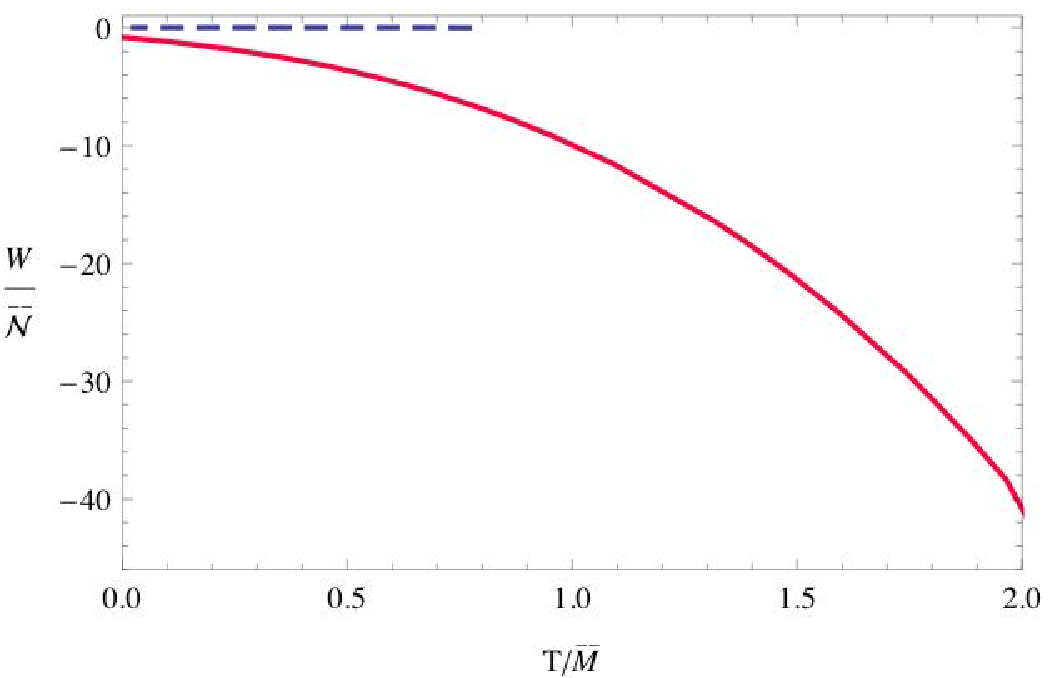}\\
 (a) $\muq/\mq = 0.8$& (b) $\muq/\mq = 2$
 \end{tabular}
\caption{Free energy versus temperature for (a) $\muq/\mq=0.8$ and
(b) $\muq/\mq=2$.  In (a), the vertical line marks the temperature of
the phase transition.} \label{freeEnVsTbig}}

By varying $\muq/\mq$ and examining the free energy, the phase
diagram in fig.~\ref{phaseDiag} was mapped out.  For $\muq/\mq<1$,
there is a phase transition corresponding to a discontinous jump
from a Minkowski to a black hole embedding.  As $\muq/\mq$ increases
towards one, the temperature of the phase transition decreases.  For
larger chemical potentials, $\muq/\mq > 1$, there is no phase
transition, with black hole embeddings thermodynamically favoured
for all temperatures.  Other thermodynamic quantities can be
computed from the free energy, \eg the entropy \mbox{$S= \left(
\partial W / \partial T \right)_{\muq}$}. Discontinuities in
thermodynamic quantities (\eg quark condensate, entropy density) at
the phase transition indicate that the phase transition is first
order. As illustrated in fig.~\ref{freeEnVsTbig}, however, the size
of the discontinuity is becoming smaller and smaller as we approach
\mbox{$\muq/\mq =1$}. In fact, recent analytic work \cite{karch3}
indicates that this is a critical point where the phase transition
becomes second order. For \mbox{$\muq/\mq <1$}, the low temperature
phase is described by Minkowski embeddings and hence $\nq=0$ below
the phase transition. After the phase transition, $\nq >0$ and it
increases as the temperature increases. Similarly, for $\muq/\mq >
1$, $\nq$ increases as the temperature increases.

Finally, we turn briefly to the thermodynamic stability of the
system. The conditions for stability can be stated in different
ways. Generally they can be phrased as the restriction that the
Hessian of the state function, here the Gibbs free energy, is
positive semi-definite. A simpler approach is to examine the
necessary (but not sufficient) conditions:
\beq \left.\frac{\partial S}{\partial T}\right|_{\muq} >0 \sac
\left.\frac{\partial n_q}{\partial \muq}\right|_{T} >0 \,, \eeq
which ensure that the grand canonical ensemble is stable against
fluctuations purely in energy or charge. In ref.~\cite{findens}, it
was stated that in the canonical ensemble the first requirement with
fixed $n_q$ (rather than fixed $\muq$) appears to be satisfied
everywhere, however, more recently ref.~\cite{korea} reported that
there are some regimes where the latter condition fails to hold. At
present, we have found no indications that the grand canonical
ensemble suffers such an energetic instability. Further,
ref.~\cite{findens} found that the second condition does not hold
for some black hole embeddings.

These unstable configurations can be found by examining
fig.~\ref{dVsmuVsT}, which illustrates $\muq$ as a function of $T$
and $\nq$ for the black hole embeddings -- note that
$\tilde{d}/m^3\propto\nq/\mq^3$ is independent of $T$. The figure
only covers a small region near the phase transition found in the
canonical ensemble. In this regime, the surface $\muq(T,\nq)$
displays an interesting `fold' and in the phase transition described
in \cite{findens}, the system jumps between configurations on the
top and bottom of this fold. One can see that the unstable
configurations discussed above correspond to the top of the fold
illustrated in fig.~\ref{dVsmuVsT}. These unstable embeddings then
play a central role in the phase transition found for the canonical
ensemble.
\FIGURE{
 \includegraphics[width=0.95 \textwidth]{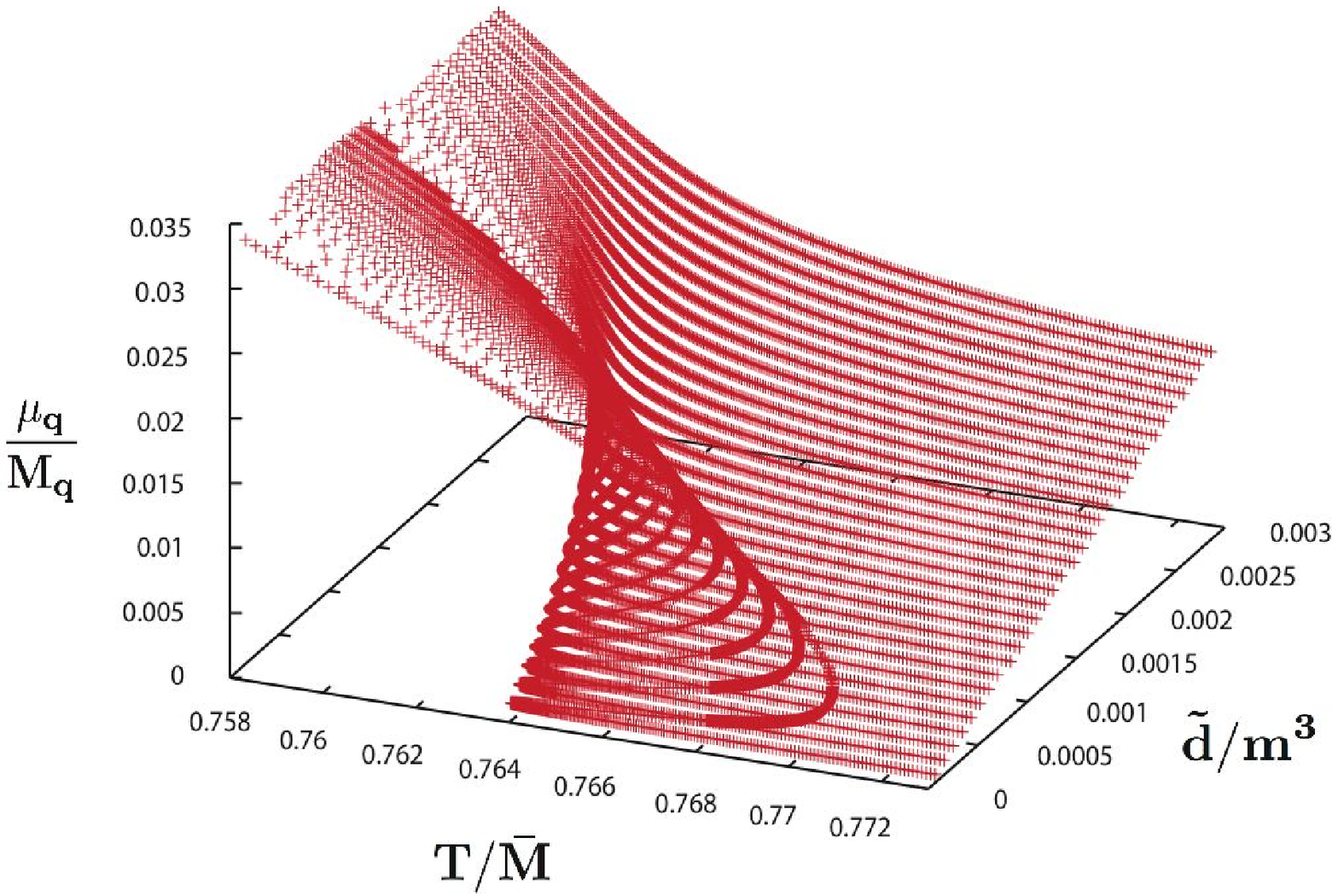}
\caption{Three-dimensional plot of the chemical potential, the
temperature and the charge density determined by black hole
embeddings.} \label{dVsmuVsT} }
In fig.~\ref{phaseDiag}, these unstable configurations appear near
the top of the multi-valued region enclosed by the blue curve and
hence they could in principle play a role in this regime for the
grand canonical ensemble as well. However, it turns out that
minimizing the Gibbs free energy always picks out either a stable
black hole or Minkowski embedding in this regime. Hence the system
never suffers from this electrical instability. This result poses an
apparent paradox though since (in the infinite volume limit, as
applies here) the canonical and grand canonical ensembles should be
equivalent \cite{thermo}, whereas here they seem to display
different behaviours. We address this issue in the discussion below.

\section{Discussion}\label{discuss}

A universal, first order, thermal phase transition in Dp/Dq systems
was identified in \cite{prl}, characterised by a jump of the
Dq-branes from a Minkowski to a black hole embedding in the Dp-brane
geometry. In the dual gauge theory, the transition is associated
with the melting or dissociation of mesons. The analysis of
ref.~\cite{prl} was extended to include a nonzero quark density in
ref.~\cite{findens}, where it was found that Minkowski embeddings
are physically inconsistent for any $\nq \neq 0$. Despite this, a
first order thermal phase transition, in this case between two black
hole embeddings, was found to occur for densities below a certain
critical density.

In this paper we have examined the thermal behaviour of the D3/D7
system as a function of the quark chemical potential $\muq$, as
opposed to the quark density $\nq$. As described in the introduction
and clearly elucidated in \cite{findens}, the restriction to black
hole embeddings only applies when $\nq$ is fixed and
nonvanishing\footnote{Refs.~\cite{korea,korea-old} attempt to
construct Minkowski embeddings with $\nq\ne0$ but we believe these
are unphysical for the reasons described in \cite{findens}.} and so
no longer applies in describing the grand canonical ensemble. In
particular then, we have found that, for $\muq < \mq$, the
low-temperature phase is described by Minkowski embeddings with $\nq
=0$. If the temperature is kept fixed and the chemical potential is
increased to a critical value, or viceversa, the system undergoes a
first order phase transition characterised by a jump  to a black
hole embedding with $\nq \neq 0$. Thus the phase transition is
characterised by the condensation of charge. The meson spectrum for
the low-temperature phase was calculated in ref.~\cite{long} and is
discrete and exhibits a mass gap.  In the high-temperature phase
there are no stable mesons and we expect a discrete spectrum of
quasinormal modes \cite{spectre,hoyos}. Hence, as with the $\muq=0$
case \cite{prl,long}, a dramatic feature characterising the
phase transition is the melting of mesons. Unlike in the $\muq=0$
case, however, there is a finite density of quarks in the
high-temperature phase, and this density increases with temperature.
Fig.~\ref{phaseDiag}a shows that the temperature of the phase
transition decreases as $\muq/\mq$ increases towards one.

For $\muq >  \mq$, there is no phase transition: In the gravity
system, there are black hole embeddings of the D7-branes
corresponding to all temperatures. Computing the free energy of
these configurations and comparing to the results for the Minkowski
embeddings reveals that the black hole embeddings are
thermodynamically preferred. It would be interesting to examine the
effect of a nonzero chemical potential on the transport properties
of the D3/D7 plasma in this regime by extending the analysis of
refs.~\cite{visco,spectre,Erdmenger:2007ap,appear}.

In this black hole phase we expect the meson spectrum to be
continuous and gapless. The true quasi-particle excitations of the
system, if any, correspond to `quasinormal modes' of the D7-brane
worldvolume fields \cite{hoyos}. In a regime in which the imaginary
part of the quasinormal frequency is small, these excitations can be
interpreted as quasi-particles in the dual quark-gluon plasma. Such
long-lived quasi-particle excitations yield particularly striking
resonances in the corresponding spectral functions. Such resonances
were found for the D3/D7-brane system with $n_q=0$ in
\cite{spectre,appear} but these were most pronounced for unstable
black hole embeddings. Hence, it would be particularly interesting
to study the spectrum of quasinormal modes \cite{spectre,hoyos} or
spectral functions \cite{spectre,appear} in detail in the present
context since, as described in \cite{findens}, we expect to find a
class of almost stable meson excitations in the regime of low
temperature and low quark density where the black hole embeddings
should be stable.

Let us expand on this point: In this regime of low temperature and
low quark density, the D7-brane connects to the event horizon with a
long narrow spike. This black hole embedding then mimics the
behaviour of a Minkowski embedding with (a gas of) fundamental strings
stretching down to the horizon \cite{findens}. Hence the quasinormal
spectrum should include long-lived modes that mimic the spectrum of
(stable) mesons on the corresponding Minkowski embedding. It is
relevant here to recall that the Minkowski phase also contains
stable unconfined quarks described by strings stretched between the
brane and the horizon. We note that these quarks also have unstable
excited states described by the quasinormal modes on the strings.
Presumably the quasinormal spectrum of the black hole embedding must
also include modes which mimic this behaviour. On the black hole
embedding, these modes describing `quark excitations' would have
support primarily on the spike connected to the horizon, in contrast
to the `mesonic' modes whose support would be primarily above the
spike. Given the standard intuition for holographic energy scales
\cite{holo-intuit}, one may expect $T<{\rm
Re}(\omega)<\mq/\sqrt{\lambda}$ for the `quark excitations'.
In contrast, the
imaginary part of the longest-lived modes is of order ${\rm
Im}(\omega)\sim\sqrt{\lambda} T^2/\mq\ll T$ in the regime considered
here \cite{seat}. Hence it is possible that these modes may
represent an independent class of quasiparticle excitations in the
quark-gluon plasma.

In the Minkowski phase the spectrum contains mesons and
quarks, and both sets of excitations are absolutely stable
in the large-$\nc$, large-$\lambda$ approximation considered here.
In view of the presence in the spectrum of stable,
quark-number-carrying states, it may seem surprising that the
physics in the Minkowski phase is completely independent of $\muq$,
in particular that the quark density vanishes in this phase. After
all, at finite temperature one expects a gas of quarks and
anti-quarks to be present, and thus a finite chemical potential to
induce a charge imbalance. The resolution of this puzzle comes from
the fact that the quark mass is of order $\mq$, whereas the maximum
temperature for which Minkowski embeddings are thermodynamically
preferred is set by the meson mass and is therefore parametrically
smaller: $T \sim \tf \sim \mq / \sqrt{\lambda}$. Thus the abundance
of quarks and anti-quarks in the thermal gas is proportional to the
Boltzman factor
\be
e^{-\mq/T} \simeq  e^{ - \sqrt{\lambda}} \,.
\ee
This is exponentially suppressed at strong coupling, and hence is
presumably invisible on the gravity side at leading order in the
large-$\lambda$, large-$\nc$ approximation under consideration.

One may also wonder whether baryons may play a role in the Minkowski
phase. From the gauge theory viewpoint, we do not expect this to be
the case, since the baryon mass scales as $\nc$ and hence the baryon
abundance at finite temperature is suppressed by $\exp \left( -\nc
\right)$. On the gravity side, at strictly zero temperature, stable
baryons are represented by D5-branes wrapping the $S^5$ part of the
geometry \cite{baryon} and connected to the D7-branes to induce an
electric flux corresponding to $\nc$ fundamental strings
\cite{us-meson}. However, our intuition is that upon introducing a
nonzero temperature,  the gravitational attraction of the black hole
causes the D5-branes to collapse and they are pulled behind the
event horizon leaving behind just a black hole embedding of the
D7-branes. We have performed some numerical investigations that seem
to support this conclusion, but our results remain preliminary.
Specifically, we have investigated the possible existence of
tear-drop configurations,\footnote{One might also consider cone-like
configurations in which the D5-brane has already partially fallen
through the event horizon. These would describe a mixed phase of
baryons and unbound quarks, if they were stable, but our intuition
is that they are also unstable.} similar to those of
\cite{callan}.\footnote{Note that the D5-brane configurations
studied in \cite{callan} do not correspond to baryons in the
four-dimensional ${\cal N} =4$ SYM theory at finite temperature. The
gravitational background in \cite{callan} corresponds to the AdS
soliton rather than an AdS black hole, so the baryons considered in
that reference are baryons in a confining phase of the gauge theory
produced by compactifying one of the dimensions.} These
configurations would end on a cusp above the horizon that could be
connected to a cuspy D7-brane configuration along a number of
fundamental strings. The tension of the strings would then overcome
the tension of the branes \cite{findens} and presumably force the
D5-branes and the D7-branes to connect to each other, resulting in a
smooth configuration. However, our preliminary results indicate that
there are no static configurations of this type. It is interesting
to imagine a dual description of the dynamical collapse of the
suggested tear-drop configurations. One would begin with a baryon as
an excited state of bound quarks which are then dispersed as free
quarks by the thermal bath of the strongly coupled gauge theory. In
any event, we re-iterate that our numerical investigations here are
only preliminary.

In the limit $T \ra 0$ or $\mq \ra \infty$, the phase transition
from a Minkowski to a black hole embedding happens exactly at $\muq
= \mq$. One can see this as arising from the fact that the D3/D7
system is supersymmetric at $T=0$. In this limit the force exerted
on the D7-branes by the background sourced by the D3-branes vanishes
and the D7-branes are not deformed \cite{us-meson}. This implies
that the constituent quark mass exactly coincides with the bare
quark mass, $\mc=\mq$. It is therefore not surprising that the phase
transition takes place when the chemical potential reaches this
mass. The constituent quark mass for $T >0$ (on Minkowski embeddings
with $\nq=0$) was calculated in \cite{long,seat}.
Fig.~\ref{comparison} shows the ratio $\mc/\mq$ (blue, dotted curve)
as a function of temperature, as well as the value of the ratio
$\muq/\mq$ (red, continuous curve) at which the phase transition
takes place.
\FIGURE{
\begin{tabular}{cc}
  \includegraphics[width=0.5 \textwidth]{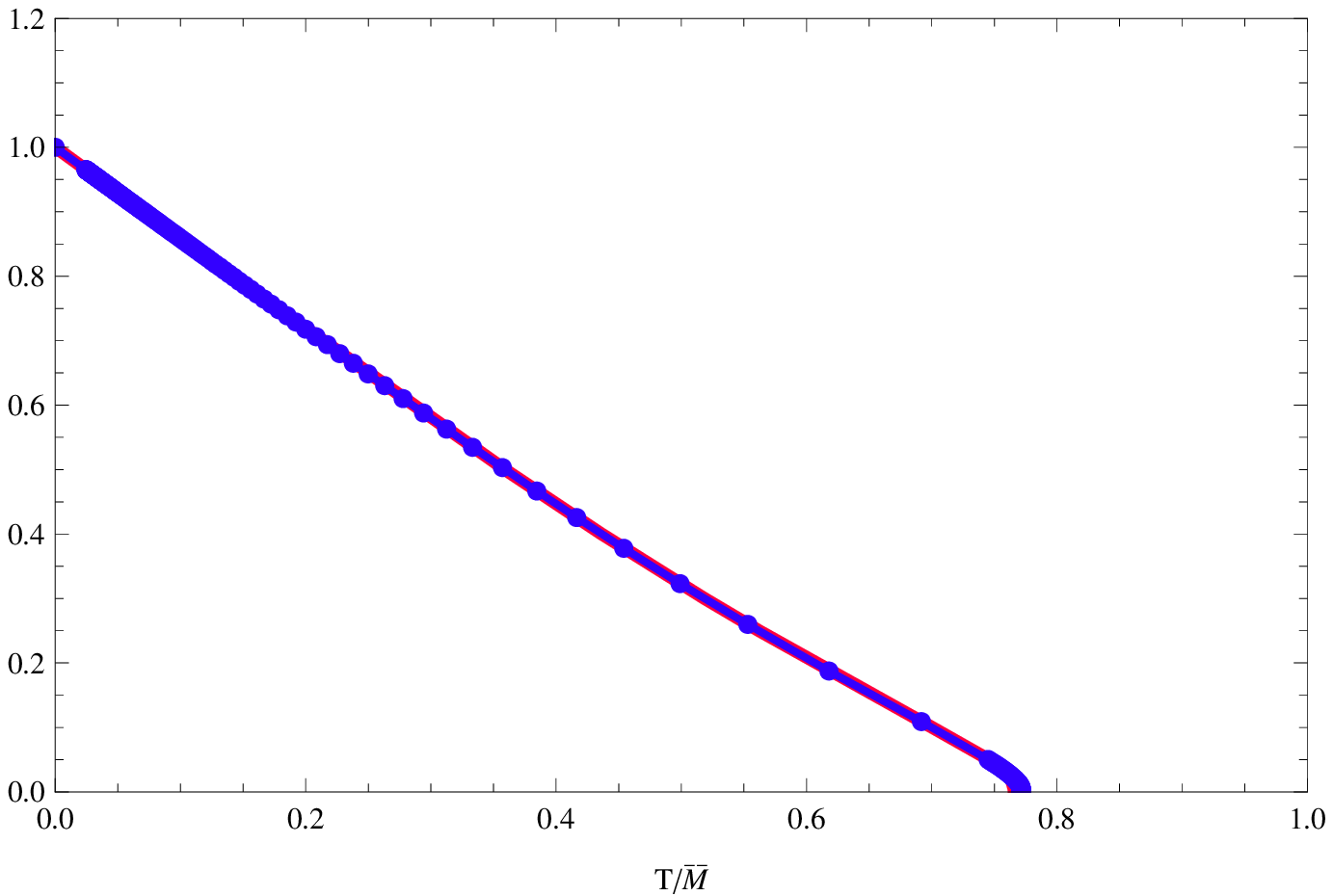}&
 \includegraphics[width=0.5 \textwidth]{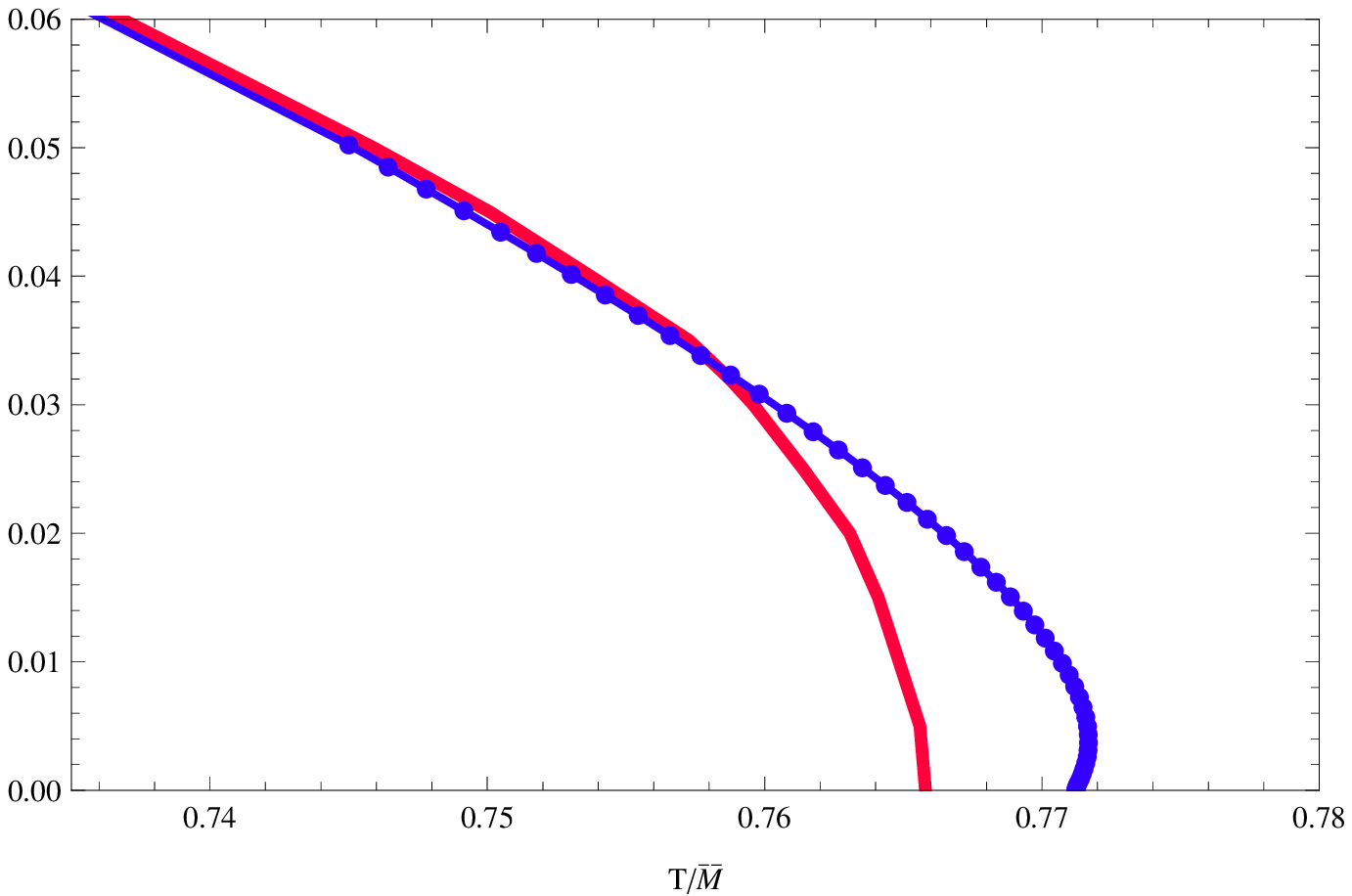}\\
 (a) & (b)
\end{tabular}
\caption{Comparison of the ratio $\mc/\mq$ (blue, dotted curve) with
the ratio $\muq/\mq$ (red, continuous curve) at which the phase
transition from a Minkowski to a black hole embedding takes place.
The two curves essentially coincide on the scale of the Figure (a).}
\label{comparison} }
It is remarkable that the two curves are almost coincident except
for relatively small values of $\muq$, \ie towards the higher
temperature end of the line of phase transitions. A similar result
was found recently in \cite{japan} for a similar holographic gauge
theory.
We note that the two curves in fig.~\ref{comparison}b begin to
diverge significantly at $T\simeq.758\mbar$, which coincides with
the temperature where the line of phase transitions enters the
multi-valued region in fig.~\ref{phaseDiag}b. This agreement seems
better than one is in principle entitled to expect. That is,
fig.~\ref{comparison} illustrates that the chemical potential
required to introduce quarks into the gluon plasma essentially
matches the energy of a `free' quark. We have put `free' in quotes
because the constituent quark mass $\mc$ does take into account the
interactions of an individual quark with the strongly coupled plasma
of adjoint fields. However, the agreement in fig.~\ref{comparison}
then indicates that the interactions between the quarks themselves
are largely negligible at the phase transition.

This result may further suggest that the density of quarks should be
small just above the phase transition. Examining fig.~\ref{djump},
we see that this statement seems to be correct in that the
dimensionless ratio $r=\tilde{d}/m^3$ is indeed much less than one.
Further we note that $r$ is largest towards the higher temperature
end of the line of phase transitions, \ie roughly in the same region
where the two curves in fig.~\ref{comparison}b deviate from one
another. However, using eqs.~(\ref{deff1}--\ref{mbar}), we find
\beq \nq = \sqrt{2}\,r\,\lambda\nf\nc\,n_\mt{crit}\qquad{\rm with}\
\ \ n_\mt{crit} \equiv
\left(\frac{\mq}{\sqrt{\lambda}}\right)^3\,.\labell{dense3} \eeq
Above we have defined $n_\mt{crit}$ keeping in mind that studies of
structure functions \cite{holycow}, as well as mapping of a quark's
disturbance of the adjoint fields \cite{riverside}, both indicate
that the size of (the glue cloud around) an individual quark is
roughly $(\mq/\sqrt{\lambda})^{-1}$ (at $T=0$). Hence we might
expect that the quarks would begin to interact significantly when
the quark density reaches $n_\mt{crit}$. Therefore, even though $r$
may be of the order $.0001$ (as at the phase transition in
fig.~\ref{djump}), eq.~\reef{dense3} indicates that $\nq \gg
n_\mt{crit}$ because we are in the limit of large $\lambda$ and
large $\nc$. Hence the good agreement between $\muq$ and $\mc$
suggests the surprising result that the quark-quark interactions
have a negligible effect even at parametrically large quark
densities.

Let us comment on the supergravity description of these results. The
constituent quark mass is defined as the energy of a test
fundamental string stretching between the tip of the D7-branes and
the horizon \cite{long,seat}. In the absence of supersymmetry (\ie
at $T>0$), this energy need not coincide with the energy of a
`spike' emanating from the D7-branes down to the horizon. This spike
in the D7-brane geometry extends uniformly in all of the gauge
theory directions. Hence this spike accounts for the effects of
the interactions of the quarks with both adjoint fields and other
quarks. In other words, this geometry, which represents the presence of a
macroscopic string density, has an energy which receives
contributions not accounted for by the test string calculation, such
as (for example) the deformation of the D7-branes.
\FIGURE{
\begin{tabular}{cc}
  \includegraphics[width=0.5 \textwidth]{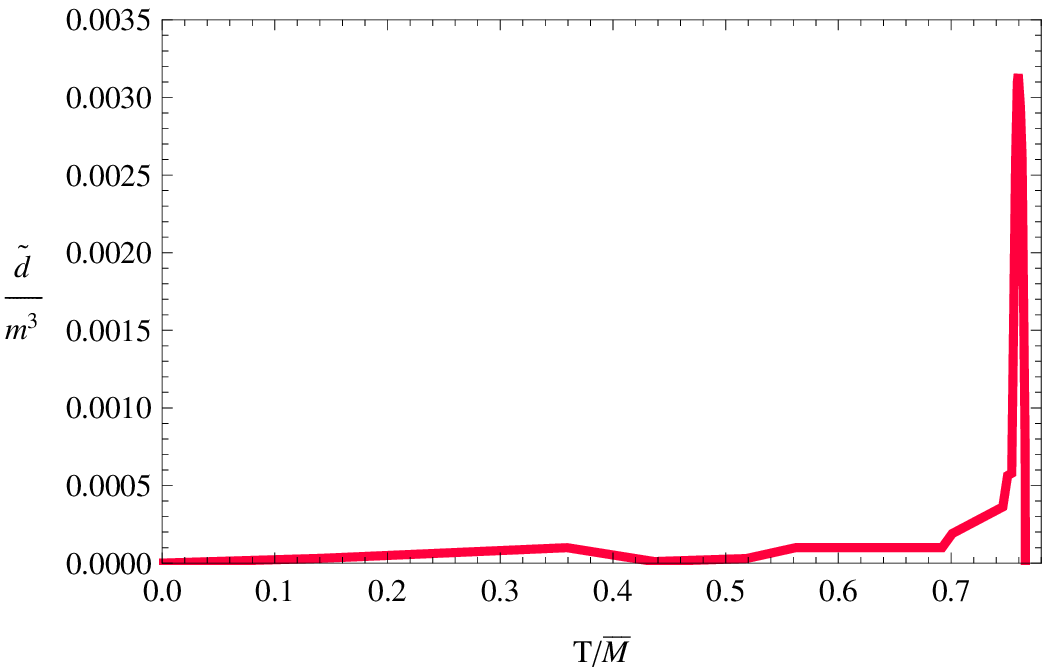}&
 \includegraphics[width=0.5 \textwidth]{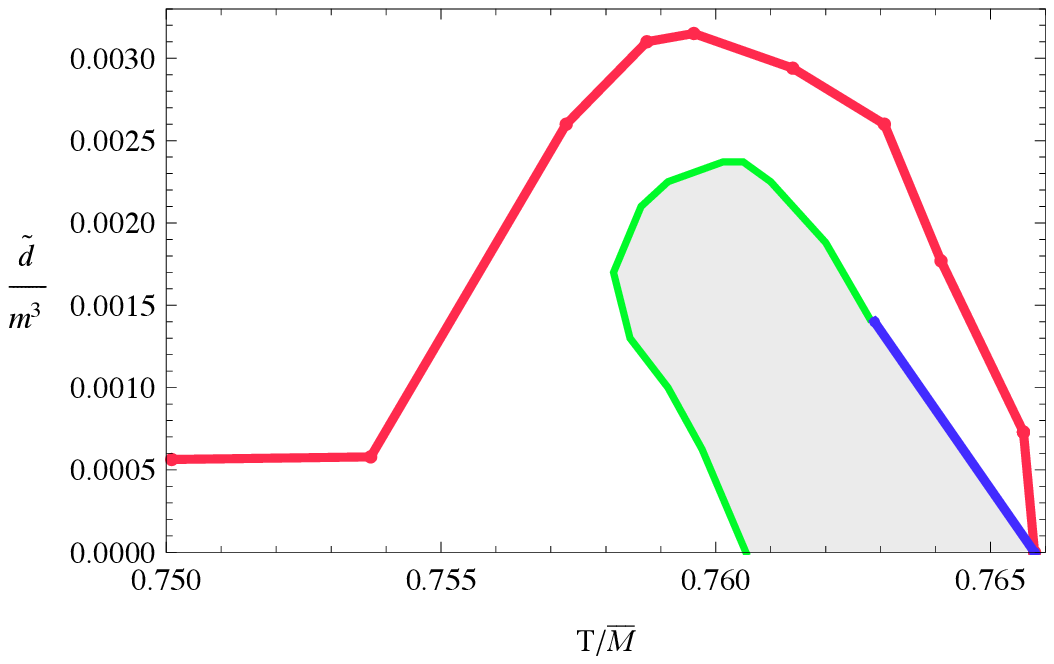}\\
 (a) & (b)
\end{tabular}
\caption{Charge density at the phase transition in the grand
canonical ensemble (red).
Figure (b) shows in blue the line of phase transitions identified in
the canonical ensemble at fixed $\nq$ \cite{findens}. Further the
shaded region enclosed by the green and blue curves corresponds to
the unstable region identified in \cite{findens}, 
in which $(\partial \muq/ \partial \nq)_T < 0$.} \label{djump} }

Before closing we would like to comment on the relation between the
present results and those presented in \cite{findens}. There for the
canonical ensemble with fixed $\nq$, we found a first order phase
transition that only existed for small charge densities and
temperatures near $T_\mt{fun}$, the critical temperature identified
at $\nq=0$. Here for the grand canonical ensemble with fixed $\muq$,
we found a first order phase transition extending from
$(\muq,T)=(0,T_\mt{fun})$ to $(\mq,0)$, as illustrated in
fig.~\ref{phaseDiag}a. The difference was further highlighted at the
end of section \ref{thermo} with the observation that certain
unstable black hole embeddings play a role in describing the phase
diagram for the canonical ensemble -- the corresponding region is
illustrated in fig.~\ref{djump}b -- but they are not physically
relevant for the present analysis of the grand canonical ensemble.
These discrepancies pose a paradox since (in the infinite volume
limit under consideration here) the canonical and grand canonical
ensembles should be equivalent \cite{thermo}.

Before resolving the inconsistency, let us elaborate on what the
equivalence stated above entails at a practical level. In the grand
canonical ensemble we fix $T$ and $\muq$. Minimizing the Gibbs free
energy then determines the charge density $\nq = \nq(T,\muq)$. In
the canonical ensemble, we fix $T$ and $\nq$ and minimize the
Helmholtz free energy. This determines the chemical potential $\muq
= \muq(T,\nq)$. However, both of these surfaces,
$(T,\muq,\nq(T,\muq))_\mt{grand}$ and
$(T,\muq(T,\nq),\nq)_\mt{canon}$, should coincide.\footnote{While
this statement is correct for practical purposes, there are, of
course, technical subtleties, \eg at the line of phase transitions
where $\partial\muq/\partial\nq=0$. Similar issues arise at $\nq=0$,
where $\partial\nq/\partial\muq=0$ in the low-temperature phase with
the present large-$\lambda$, large-$\nc$ approximation.} That is,
a given $T$ and $\muq$ in the grand canonical ensemble determine a
particular value of $\nq$. Then with this value of $\nq$ and the
same temperature $T$, thermal equilibrium in the canonical ensemble
will yield the same value of $\muq$ as before.

Now the resolution of our inconsistency can be found by examining
fig.~\ref{djump} which illustrates the quark density on the black
hole embedding along the line of phase transitions in
fig.~\ref{phaseDiag}. Recall that on the other side of the phase
transition, the Minkowski embeddings have $\nq=0$ (or
$\tilde{d}=0$). Hence if we map each configuration in the phase
plane in fig.~\ref{phaseDiag}a to the $(\nq,T)$ plane as in
fig.~\ref{djump}, we find a gap between the $\nq=0$ axis and the red
curve illustrated in the second figure. Hence we are left to explain
how the system accesses these values of $\nq$ and $T$. Here we can
use the intuition that we might derive from, say, the liquid to gas
transition of ordinary water. That is, we expect that this
intermediate regime is filled in by an inhomogeneous mixture where
the two phases (with the same chemical potential) coexist.
Now the key point, shown in fig.~\ref{djump}b, is that the phase
transition identified for the canonical ensemble in \cite{findens},
as well as the region where unstable embeddings were favoured in
this ensemble, lies entirely within this intermediate regime where
we have just argued that the system should be inhomogeneous. Since
the analysis of \cite{findens} was restricted to only consider the
homogeneous configurations, it misidentified the correct ground
state of the system in the small region around the phase transition.

Of course this is not a surprise since, as discussed in
\cite{findens}, the electrical instability of the configurations
below the phase transition there already pointed towards an
inhomogeneous phase playing a role in the correct phase diagram. The
present discussion gives a clearer picture of the nature of this
inhomogenous phase and where it is relevant. Let us add that this
clarification also highlights a shortcoming of the holographic
explorations of the phase diagram of strongly coupled gauge theories
here and, for example, in
\cite{findens,korea,japan,karch3,korea-old,ss-chemical}.
That is, they rely on constructing simple and
highly symmetric configurations of D-branes and supergravity fields
-- in particular, configurations describing homogeneous phases of
the gauge theory. These symmetries are imposed by hand and,
since it is difficult to account for deviations towards less
symmetric configurations, the latter are typically ignored. As
highlighted by the present discussion, this
simplistic approach may fail even in relatively simple
circumstances, and so one must be cautious in extracting
conclusions based on highly symmetric configurations.

Let us finish with some comments on the possible comparison of the
phase diagram obtained here with that of QCD. In addition to the
usual caveats at zero density, there are two additional caveats that
are particularly relevant at finite chemical potential. The first
one is the fact that baryon number in QCD is only carried by
fermionic fields, whereas in the type of systems we have studied it
is also carried by scalar fields in the fundamental representation.
This feature is absent in the D4/D8/$\bar{\mbox{D8}}$ system of
Sakai and Sugimoto \cite{ss}, whose spectrum in the fundamental
representation consists only of fermions. Interesting analyses of
this system at finite baryon number chemical potential have recently
appeared \cite{ss-chemical}, and further studies will likely be a
fruitful avenue for the future. The second caveat is the fact that
the very existence of  many of the expected phases of QCD at finite
chemical potential depend crucially on the fact that $\nc=3$ (see
\eg \cite{cond,Stephanov:2007fk}). It is therefore unclear to what
extent the large-$\nc$ limit may provide a useful approximation at
finite chemical potential; some optimistic arguments, however, can
be found in \cite{optimistic}. In any case, given the lack of other
analytical methods and the difficulties of simulating QCD at finite
chemical potential on the lattice, we think it is interesting to
explore possible insights coming from the holographic duals of
large-$\nc$ gauge theories.

\acknowledgments It is a pleasure to thank Jorge Casalderrey-Solana,
Jaume Gomis, Volker Koch, Jorgen Randrup and Andrei Starinets for useful correspondence and conversations. Research at the Perimeter Institute is supported in
part by funds from NSERC of Canada and MEDT of Ontario. We also
acknowledge support from NSF grant PHY-0555669 (DM), an NSERC
Discovery grant (RCM), a JSPS Research Fellowship for Young
Scientists (SM), an NSERC Canada Graduate Scholarship (RMT) and
funding from the Canadian Institute for Advanced Research (RCM,RMT).
RCM would also thank the Kavli Institute for Theoretical Physics for
hospitality at the beginning of this project. Research at the KITP
was supported in part by the NSF under Grant No. PHY05-51164.

\end{document}